\documentclass[RNAAS]{aastex62}

\begin{document}

\newcommand{\kms}{\ensuremath{\mathrm{km}\,\mathrm{s}^{-1}}}
\newcommand{\galunits}{\ensuremath{\mathrm{km}\,\mathrm{s}^{-1}\,\mathrm{kpc}^{-1}}}
\newcommand{\galacc}{\ensuremath{\mathrm{km}^2\,\mathrm{s}^{-2}\,\mathrm{kpc}^{-1}}}
\newcommand{\MLsun}{\ensuremath{\mathrm{M}_{\sun}/\mathrm{L}_{\sun}}}
\newcommand{\Lsun}{\ensuremath{\mathrm{L}_{\sun}}}
\newcommand{\Msun}{\ensuremath{\mathrm{M}_{\sun}}}
\newcommand{\Ha}{\ensuremath{\mathrm{H}\alpha}}
\newcommand{\SFR}{\ensuremath{\mathit{SFR}}}
\newcommand{\aveSFR}{\ensuremath{\langle \mathit{SFR} \rangle}}
\newcommand{\sfrate}{\ensuremath{\mathrm{M}_{\sun}\,\mathrm{yr}^{-1}}}
\newcommand{\Aunits}{\ensuremath{\mathrm{M}_{\sun}\,\mathrm{km}^{-4}\,\mathrm{s}^{4}}}
\newcommand{\surfdens}{\ensuremath{\mathrm{M}_{\sun}\,\mathrm{pc}^{-2}}}
\newcommand{\voldens}{\ensuremath{\mathrm{M}_{\sun}\,\mathrm{pc}^{-3}}}
\newcommand{\gevcc}{\ensuremath{\mathrm{GeV}\,\mathrm{cm}^{-3}}}
\newcommand{\etal}{et al.}
\newcommand{\LCDM}{$\Lambda$CDM}
\newcommand{\ML}{\ensuremath{\Upsilon_*}}
\newcommand{\Mst}{\ensuremath{M_*}}
\newcommand{\Mg}{\ensuremath{M_g}}
\newcommand{\Mb}{\ensuremath{M_b}}

\title{Strong Hydrogen Absorption at Cosmic Dawn: \\ the Signature of a Baryonic Universe}

\author{Stacy S. McGaugh}
\affil{Department of Astronomy, Case Western Reserve University, Cleveland, OH 44106}



\section{Introduction}

\citet{EDGES} recently reported the detection of redshifted 21cm absorption at $z \approx 17$. 
The observed strength of this signal is anomalously strong for \LCDM.
Here I show that this signal is expected in a purely baryonic universe \citep{M1999fb}.

\section{The cosmic baryon fraction}

In \LCDM, most of the mass is in the form of non-baryonic cold dark matter. This is a radical hypothesis that remains to be confirmed
by a laboratory detection. Alternatively, it is conceivable that the missing mass problem indicates a breakdown of known
dynamical laws: the radical hypothesis of modified gravity. Measurements 
sensitive to the baryon fraction ($f_b = \Omega_b/\Omega_m$) of the universe provide a distinguishing test.
If non-baryonic dark matter is indeed the dominant form of mass in the universe, then the baryon fraction is $f_b = 0.16$ \citep{Planck14}.
If CDM does not exist, then $f_b = 1$. 

The detection of 21cm absorption reported by \citet{EDGES} provides a test for the baryon fraction. 
This signal arises from the decoupling of the spin temperature $T_S$ of the 21cm line from the radiation temperature,
causing the predominantly neutral IGM to be seen in absorption against the CMB \citep{ZFH,PL12}.
The amount of absorption depends on the baryon density, with little sensitivity to other cosmological parameters or to 
the necessary astrophysical details. 

This is a problem of radiative transfer in the early universe. It depends
on known atomic physics, not whatever new physics causes the mass discrepancy. 
Writing eq.~2 of \citet{ZFH} in terms of the baryon fraction, the amplitude of absorption is
\begin{equation}
T_{21}(z) = (20\; \mathrm{mK})\, x_{\mathrm{HI}} \left[ (1+z) f_b \left( \frac{\Omega_b h^2}{0.02} \right) \right]^{1/2} \left(1- \frac{T_{\mathrm{CMB}}}{T_S} \right)
\label{eq:T21}
\end{equation}
where $x_{\mathrm{HI}}$ is the neutral hydrogen fraction. The absolute baryon density is known from BBN \citep[][]{BBN},
the CMB temperature declines as the universe expands, the kinetic temperature of the gas declines faster,
and the spin temperature is bounded between the two \citep[$T_K \le T_S \le T_{CMB}$;][]{CFBL}.
The baryon fraction is the only variable that can increase the the absorption within these bounds.

The maximum possible signal occurs when $x_{\mathrm{HI}} = 1$ and $T_S = T_K$. 
At $z=17$, $T_{\mathrm{CMB}}/T_K \approx 8.1$ \citep[Fig.~5 of ][]{CFBL}. Equation \ref{eq:T21} then predicts a maximum absorption of 
\begin{eqnarray}
T_{21,\;\mathrm{max}} = -0.24\;\mathrm{K}\;\mathrm{for}\;f_b = 0.16; \label{eq:maxTLCDM} \\
T_{21,\;\mathrm{max}} = -0.60\;\mathrm{K}\;\mathrm{for}\;f_b = 1.\phn\phn\phn \label{eq:maxTfb1}
\end{eqnarray}
The observed value is $T_{21} \approx -0.5$ K \citep{EDGES}, and appears from their Fig.\ 2 to be closer to $-0.55$ K.
Such a strong signal is expected in a baryonic universe, but is unobtainable\footnote{\citet{Barkingmad} engages in
special pleading to artificially decrease $T_K$. Allowing $T_S = T_K \rightarrow 0$ facilitates arbitrary signal strength.} in \LCDM.

\begin{figure}
\plotone{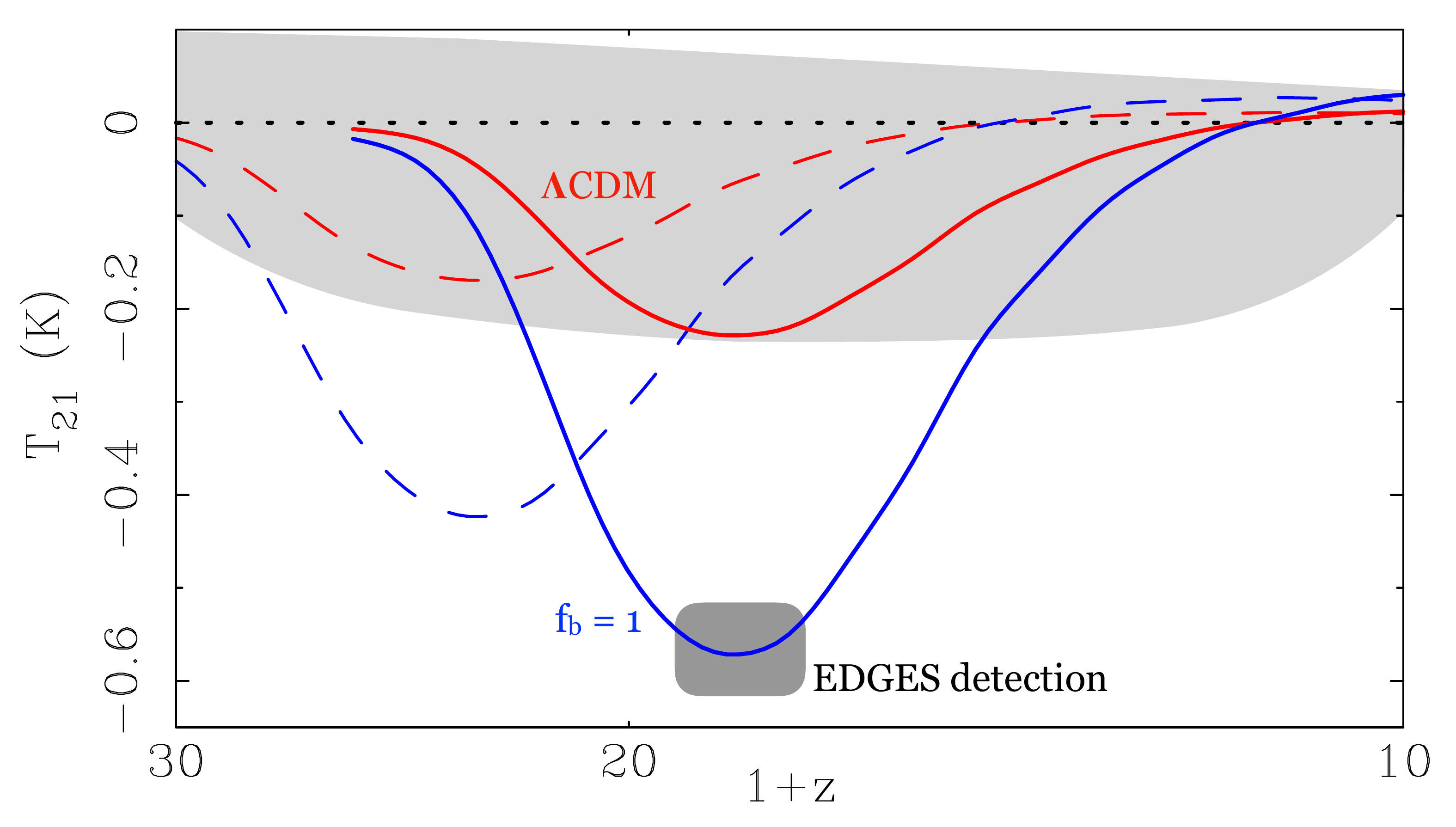}
\caption{The predicted 21cm absorption signal as a function of redshift. Red lines are \LCDM\ models from \citet{CFBL}.
The dashed red line is their standard model; the solid red line represents the maximum absorption attained at $z \approx 17$.
The light gray area illustrates the range of parameter space occupied for a wide range of astrophysical parameters 
\citep[see Fig.~1 of][]{CFBL}. Blue lines represent baryonic models, which have deeper absorption by the factor $f_b^{-1/2} = 2.5$ (eq.\ \ref{eq:T21}), 
consistent with the detection by \citet[dark gray]{EDGES}.
}
\label{fig:Tb}
\end{figure}

Fig.\ \ref{fig:Tb} shows models for the 21cm absorption adopted from \citet{CFBL}, who make a thorough exploration
of astrophysical effects. These affect the redshift at which the absorption feature appears, and can modulate its amplitude
somewhat. A stronger signal than the limit given in eq.\ \ref{eq:maxTLCDM} is profoundly unnatural in \LCDM.

\section{Further Predictions}

A review of dark matter and modified gravity is beyond the scope of this note \citep[see][]{MdB98b,SM02,FM12,CJP}. 
It suffices to say that there are remarkable genuine successes and apparently insurmountable hurdles for both approaches \citep[see][]{MerrittPhilo}.
Here I make a few additional predictions for a baryonic universe:
\begin{enumerate}
\item Strong 21cm absorption will also be observed during the dark ages ($z > 30$).
\item The 21cm power spectrum will show pronounced baryonic features. 
\item Large galaxies and the cosmic web emerge earlier than anticipated in \LCDM\ \citep{Sanders98,CJP}.
\end{enumerate}
The first two predictions stem simply from a universe made of baryons.
Only the third prediction is model-specific; some hints of large early structures already exist \citep[e.g.,][]{impossiblyearly,CCPCII}.
This implies that structure grows nonlinearly, erasing baryonic features at late times through mode mixing \citep{M1999fb}.

\end{document}